\documentclass[twocolumn,prl,aps,10pt,footinbib]{revtex4-1}
\usepackage{graphicx}
\usepackage[utf8x]{inputenc}
\usepackage{amsmath}
\usepackage{lineno}

\begin{document}

\title{Manifestations of the electron-phonon interaction range in  angle resolved photoemission spectra}

\author{J. Krsnik$^{1}$, V. N. Strocov$^2$, N. Nagaosa$^{3,4}$,  
O. S. Bari\v{s}i\'{c}$^1$, Z. Rukelj$^{5}$, S. M. Yakubenya$^{6}$, 
and  A. S. Mishchenko$^{3,6}$}
 
\affiliation{
$^1$Institute of Physics, Bijeni\v{c}ka 46, 10000 Zagreb, Croatia\\
$^2$Swiss Light Source, Paul Scherrer Institute, Villigen CH-5232, Switzerland\\
$^3$RIKEN Center for Emergent Matter Science (CEMS), Wako, Saitama 351-0198, Japan\\
$^4$Department of Applied Physics, The University of Tokyo 7-3-1 Hongo, Bunkyo-ku, Tokyo 113-8656, Japan\\
$^5$Department of Physics, Faculty of Science, University of Zagreb, Bijeni\v cka c. 32, 
HR-10000 Zagreb, Croatia\\
$^6$NRC ``{\it Kurchatov Institute}'', 123182 Moscow, Russia
}

\begin{abstract}
Numerous angle resolved photoemission spectroscopy (ARPES) studies of a wide class of low-density  metallic systems, ranging from doped transition metal oxides to quasi two-dimensional interfaces between insulators, exhibit phonon sidebands below the quasi-particle peak as a unique hallmark of  polaronic correlations. 
Here, we single out properties of ARPES spectra that can provide a robust estimate of the effective range (screening length) of the electron-phonon interaction, regardless of the limited experimental resolution, dimensionality and particular features of the electronic structure, facilitating a general methodology for an analysis of a whole class of materials. 
\end{abstract}

\maketitle

{\it Introduction.} It has been well established, both experimentally and theoretically, that for low concentrations of itinerant charge carriers the electron-phonon interaction (EPI) produces phonon sidebands, appearing as satellites in ARPES spectra below the quasi-particle peak (QP) \cite{Alexandrov}. It has been argued \cite{Asensio2018,ExponentTheory2016} that energy and momentum intensity distributions of these satellites depend on the spatial range of the EPI, the unscreened polar Fr\" ohlich interaction having the longest and the Holstein on-site interaction the shortest range. However, despite extensive ARPES data available
\cite{TiO2MoserBarisic2013,ChenAsensio2015,TiO2-SmallDoping-RIXS2015,ZnOARPES_2016,Cancellieri2016,Wang2016,TiO2-SmallDoping2017,Zhang2017,Riley2018,husanu2020}, we are not aware of any systematic theoretical study which would consider the dependence of these very unique manifestations of the EPI on the screening length. We can mention only a study \cite{Asensio2018} of the screened Fr\" ohlich interaction for few particular cases in specific materials and the theoretical study \cite{ExponentTheory2016} of the EPI with a hypothetical forward scattering. Hence, an accurate theoretical description of ARPES spectra in dependence on the screening length is of primary importance for a knowledge of the effective interaction that governs polaronic correlations in a plethora of materials \cite{TiO2MoserBarisic2013,ChenAsensio2015,TiO2-SmallDoping-RIXS2015,ZnOARPES_2016,Cancellieri2016,TiO2-SmallDoping2017,Wang2016,TiO2-SmallDoping2017,Zhang2017,Riley2018,husanu2020}, exhibiting metal-insulator transitions, colossal magnetoresistance, commensurate-incommensurate transition, quantum Hall effect, etc.

Recently, first principle calculations \cite{Giustino2017} have been used, considering the leading-order expansion of the electron self-energy \cite{Antonius2015,P6} and the cumulant expansion \cite{Story2014,TiO2-SmallDoping2017,Riley2018}, to simulate measured ARPES spectra of some specific polaronic materials. In their implementation of the cumulant expansion, the latter works are restricted to self-energy diagrams with electron Green’s functions propagating in one direction of time only \cite{Gumhalter2016}, giving approximate results for finite electron densities, including the lowest-order case. For the higher order corrections, additional approximations are introduced by treating higher phonon processes as uncorrelated \cite{Dunn1975}. On the other hand, our calculations are based on a direct evaluation of self-energy diagrams, considering the exact momentum and frequency dependence of the leading corrections to the phonon sidebands, including the exact leading (second-order) vertex correction to the second sideband. Instead of focusing on some particular system, our analysis identifies different behaviors across all ranges of EPI and provides a general procedure for the estimation of the EPI range from experimental data, prior to material specific calculations.
   
{\it Modeling.} We consider the standard EPI model  for a $D$-dimensional lattice, describing the interaction between bare electrons in a band, $\hat{H}_{\mbox{\scriptsize el}} = \sum_{{\bf k}}  \epsilon_{\bf k} a_{{\bf k}}^{\dagger} a_{{\bf k}}$, and dispersionless optical phonons with frequency $\omega_0$, 
$\hat{H}_{\mbox{\scriptsize ph}} =
\omega_0\sum_{\bf q} b_{{\bf q}}^{\dagger} b_{{\bf q}}$. The screened Fr\" ohlich interaction is given by,

\begin{equation}
\hat{H}_{\mbox{\scriptsize el-ph}} = 
\sum\limits_{{\bf k},{\bf q}} V^{d}_{\bf q}
\;a_{{\bf k}+{\bf q}}^{\dagger} a_{{\bf k}} 
(b^{\dagger}_{-{\bf q}}+ b_{{\bf q}})\;,
\label{H_int}
\end{equation}

\noindent where $\left| V^{d}_{\bf q} \right|^2 = a_d / \left( {|\bf q}|^{d-1} + q_{TF}^{d-1} \right)$, 
with $a_{d=3} = 2 \sqrt{2} \pi \alpha$ for three- (3D) and $a_{d=2} = \sqrt{2} \pi \alpha$ 
for two-dimensional (2D) systems
\cite{2D-Screening-SternHoward1967,2D-ScreenningPortnoi1997,2D-ScreeningOuerdane2011,Peeteres-2D-Frohlich1986,DevreeseAlexandrovRepProgPhys2009}, and $\alpha$ characterizing the strength of the EPI \cite{Peeteres-2D-Frohlich1986}. We assume a static screening characterized by the Thomas-Fermi wave number $q_{TF}$. $r_{TF} = \pi/q_{TF}$ roughly gives the screening range in the lattice constant units, with $q_{TF}\rightarrow0$ ($q_{TF}\rightarrow\infty$) for the unscreened Fr\" ohlich (Holstein) EPI.

The spectral function $A({\bf k},\omega)$, measured by ARPES for $\omega<0$ (electron removal processes), is determined by the electron Green function $G({\bf k},\omega)$, $A({\bf k},\omega) = |\mbox{Im}G({\bf k},\omega)|/\pi$. In terms of the 
self-energy $\Sigma({\bf k},\omega)$, appearing due to the EPI in Eq.~(\ref{H_int}), it is given by,

\begin{equation}
A({\bf k},\omega) = 
\frac{|\mbox{Im}\Sigma({\bf k},\omega)|}
{\left[ \omega - \xi_{\bf k} - \mbox{Re}\Sigma({\bf k},\omega) \right]^2 
+ \left[ \mbox{Im}\Sigma({\bf k},\omega)\right]^2} \; ,
\label{SF}
\end{equation}

\noindent with $\xi_{\bf k} = \epsilon_{\bf k}-\mu<0$, the hole energy measured from the Fermi level $\mu$. For a weak EPI, the self-energy contributions in the denominator of Eq.~(\ref{SF}) may be neglected. For a stronger EPI, the real part of the self-energy may be absorbed \cite{AandR} into a renormalized dispersion $\widetilde{\epsilon}_{\bf k}$, $\widetilde{\epsilon}_{\bf k} = \epsilon_{\bf k} + \mbox{Re}\Sigma({\bf k},\omega)$. This facilitates experimental analysis, because $\widetilde{\epsilon}_{\bf k}$ is a measured quantity. We neglect the imaginary part $\mbox{Im}\Sigma({\bf k},\omega)$ in the denominator of Eq.~(\ref{SF}) since it barely affects the ${\bf k}$ and $\omega$ dependence of the spectral density $A({\bf k},\omega) $.

Our attention is on recent ARPES studies of polaronic materials, when only a small part of the conduction band $\epsilon_{\bf k}$ is filled. The EPI leads to a complex structure of $A({\bf k},\omega)$, where in addition to the QP peak one observes phonon sidebands, with the $n$-th sideband being shifted downward from the Fermi level by $n\omega_0$ \cite{Asensio2018,ExponentTheory2016}. 
Since the ARPES seldom show more than two sidebands \cite{TiO2-SmallDoping-RIXS2015,Cancellieri2016,TiO2-SmallDoping2017,Riley2018}, we concentrate our analysis on these.  In particular, we consider the $\mu < \omega_0$ case that ensures a pattern of spectrally separated phonon sidebands, experimentally reported for many different systems 
\cite{TiO2MoserBarisic2013,ChenAsensio2015,TiO2-SmallDoping-RIXS2015,ZnOARPES_2016,Cancellieri2016,TiO2-SmallDoping2017,Zhang2017,Riley2018}.  Within the zero-temperature diagrammatic expansion, the first order contribution in $\alpha$ to $\mbox{Im}\Sigma$ \cite{Mahan},

\begin{eqnarray}
 \mbox{Im}\Sigma^{(1)}({\bf k},\omega)& =&
 \pi \sum^{BZ}_{\bf q} \left| V^{d}_{\bf q} \right|^2 \nonumber\\
&\times& \Theta(-\xi_{\bf k+q})   
 \delta \left( \omega - \xi_{\bf k+q} + \omega_0 \right) \; , 
\label{alpha1}
\end{eqnarray}

\noindent corresponds to the leading contribution to the first phonon sideband, restricted to the frequency window $[-\omega_0-\mu, - \omega_0]$. The second sideband is found in the window $[-2\omega_0-\mu, - 2\omega_0]$, with the leading contribution given by second-order $\alpha^2$ terms derived here (for details, see the Supplemental Material \cite{Supplement}),

\begin{eqnarray}
& \mbox{Im}\Sigma^{(2)}({\bf k},\omega) = \pi \sum\limits^{BZ}_{\bf q} 
\Theta(-\xi_{\bf q})\delta(\omega -  \xi_{\bf  q} + 2 \omega_0)  & \nonumber \\ 
&\times   
\sum\limits^{BZ}_{\bf q'}
 \dfrac{\left| V^{d}_{\bf q'} \right|^2  \left| V^{d}_{\bf q-k-q'} \right|^2}
{(\xi_{\bf  q}-\xi_{\bf  k+q'}-\omega_0)}
  \nonumber \\
\times& 
\left[
\left( \xi_{\bf  q} - \xi_{\bf k+q'} - \omega_0 \right)^{-1} +
\left( \xi_{\bf  q} - \xi_{\bf q-q'} - \omega_0 \right)^{-1}
\right] \; .
& 
\label{alpha2}
\end{eqnarray}

\noindent The first term in the last line of Eq.~(\ref{alpha2}) is given by the non-crossing two-phonon diagram, whereas the crossing diagram with the leading vertex correction gives the second term. 

We set $\omega_0=1$ and consider the dispersion for a cubic lattice in 3D/2D, $\epsilon_{\bf k} = 2t \sum_{i=1}^{D} \left( 1-\cos(k_i) \right)$,  where $t$ is the nearest neighbor hopping. With the lattice constant $a=1$, the effective mass of the non-interacting electron at the bottom of the band is $m_0=1/2t$. To search for general properties of ARPES sidebands components in various systems, we use two very different sets of parameters for the broad and the narrow electron band: S1 (S2) denotes $\mu=0.5$ ($\mu=0.1$) and $t=1$ ($t=1/24$), with mass $m_0=0.5$ ($m_0=12$). In both these cases, only a small fraction of the lowest band states is occupied.

\begin{figure}[tbp]
\begin{center}
\includegraphics[scale=0.18]{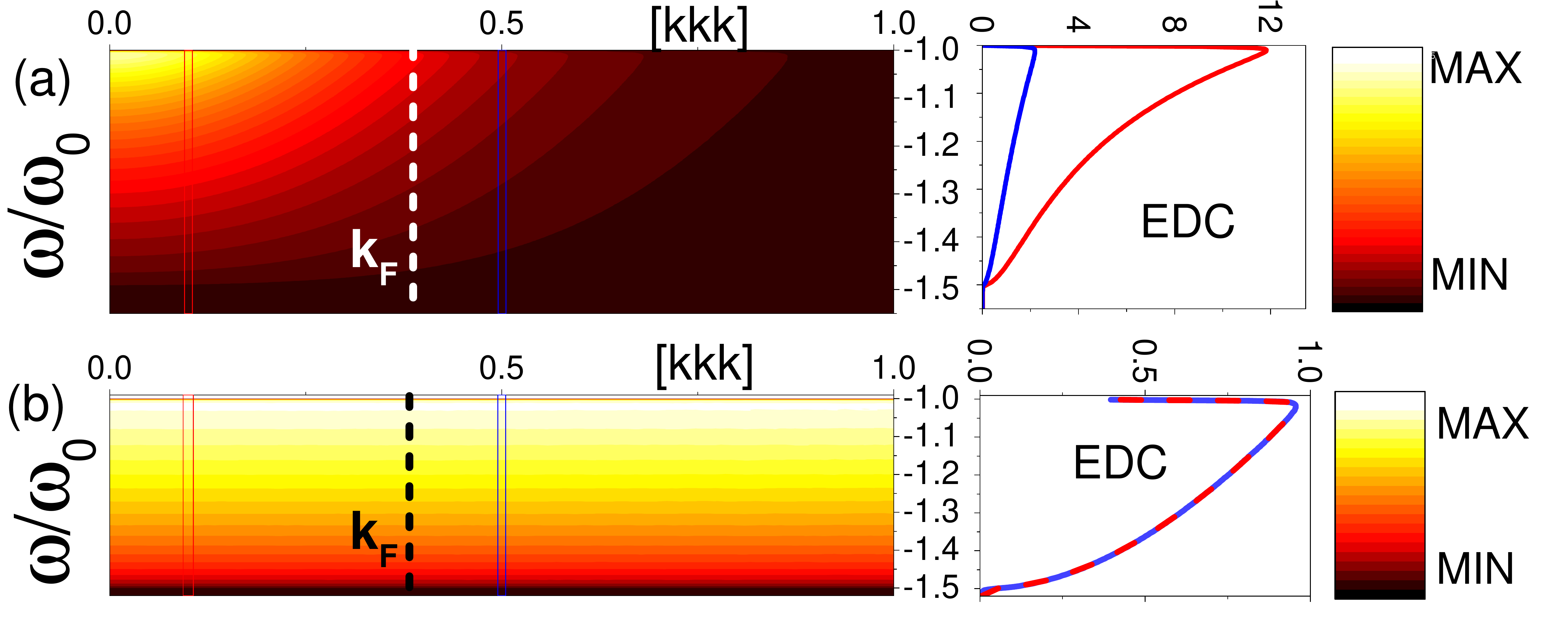}
\caption{Contour plot of the (a) ARPES component 
$A^{(1)}({\bf k},\omega)$ and 
(b) $\mbox{Im}\Sigma^{(1)}({\bf k},\omega)$ for the first sideband for the 3D parameter set S1 and strong screening $r_{TF}=0.1$. Right panels show EDCs for the corresponding cross sections in left panels. Note, two EDCs for $\mbox{Im}\Sigma$ coincide.}
\label{fig:fig1}
\end{center}
\end{figure}

{\it Phonon sidebands.} A typical example of the ARPES intensity and of the corresponding $\mbox{Im}\Sigma$ intensity for the first sideband and strong screening is shown in Fig.~\ref{fig:fig1}. In experimental works, it is often concluded that the higher intensity accumulated within the Fermi surface (FS), $k\leq k_F\ll\pi$, identifies the weakly screened Fr\"ohlich interaction. However, a more careful inspection of Eq.~(\ref{SF}) leads to a different interpretation. $A^{(1)}({\bf k},\omega)$ in Fig.~\ref{fig:fig1}a exhibits a strong momentum dependence, due to $(\omega-\tilde\xi_{\bf k})^{-2}$ in the denominator in Eq.~(\ref{SF}), which highlights the area within the FS. In fact, the latter easily camouflages the fact that the EPI might be over-screened. Indeed, as shown in Fig.~\ref{fig:fig1}b, for strong screening $\mbox{Im}\Sigma^{(1)}({\bf k},\omega)$ is momentum independent and uniformly spread over the whole Brillouin zone (BZ): in the right panel of Fig.~\ref{fig:fig1}b, the two energy distribution curves (EDCs) for $\mbox{Im}\Sigma^{(1)}({\bf k},\omega)$ are almost indistinguishable. The $\omega$-dependence is determined solely by the density of occupied electron states at the bottom of the band. This may easily be derived from Eq.~(\ref{alpha1}), assuming a momentum independent interaction $V^{d}_{\bf q}$. Different behaviors of $A^{(1)}({\bf k},\omega)$ and $\mbox{Im}\Sigma^{(1)}({\bf k},\omega)$ in Fig.~\ref{fig:fig1} explain the universality of the experimentally found confinement of the intensity of the ARPES sidebands in the momentum-space region near the band minimum \cite{TiO2MoserBarisic2013,ChenAsensio2015,TiO2-SmallDoping-RIXS2015,ZnOARPES_2016,Cancellieri2016,TiO2-SmallDoping2017,Riley2018}. However, it emphasizes as well that very different confinements may characterize the ARPES spectrum and the corresponding $\mbox{Im}\Sigma$ intensity, rendering $\mbox{Im}\Sigma$ as a quantity that is far more suitable for estimation of the range of EPI.

\begin{figure}[tbp]
	\begin{center}
		\includegraphics[scale=0.4]{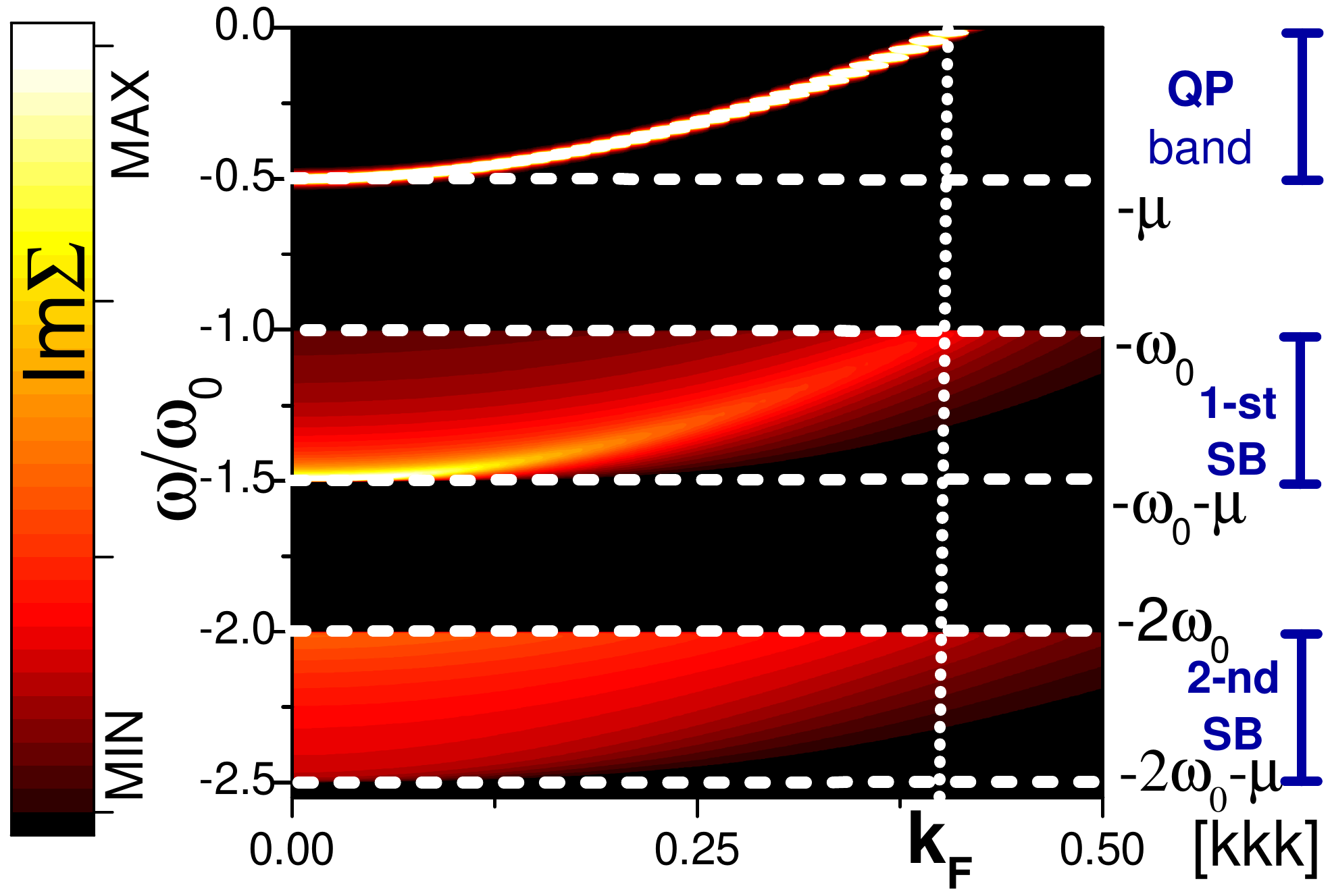}
		\caption{Contour plots of $\mbox{Im}\Sigma$ for the QP band, the first and the second 
		phonon sideband for the 3D parameter set S1 and for weak screening $r_{TF}=100$.}
		\label{fig:fig3}
	\end{center}
\end{figure}

\begin{figure}[tbp]
	\begin{center}
		\includegraphics[scale=0.45]{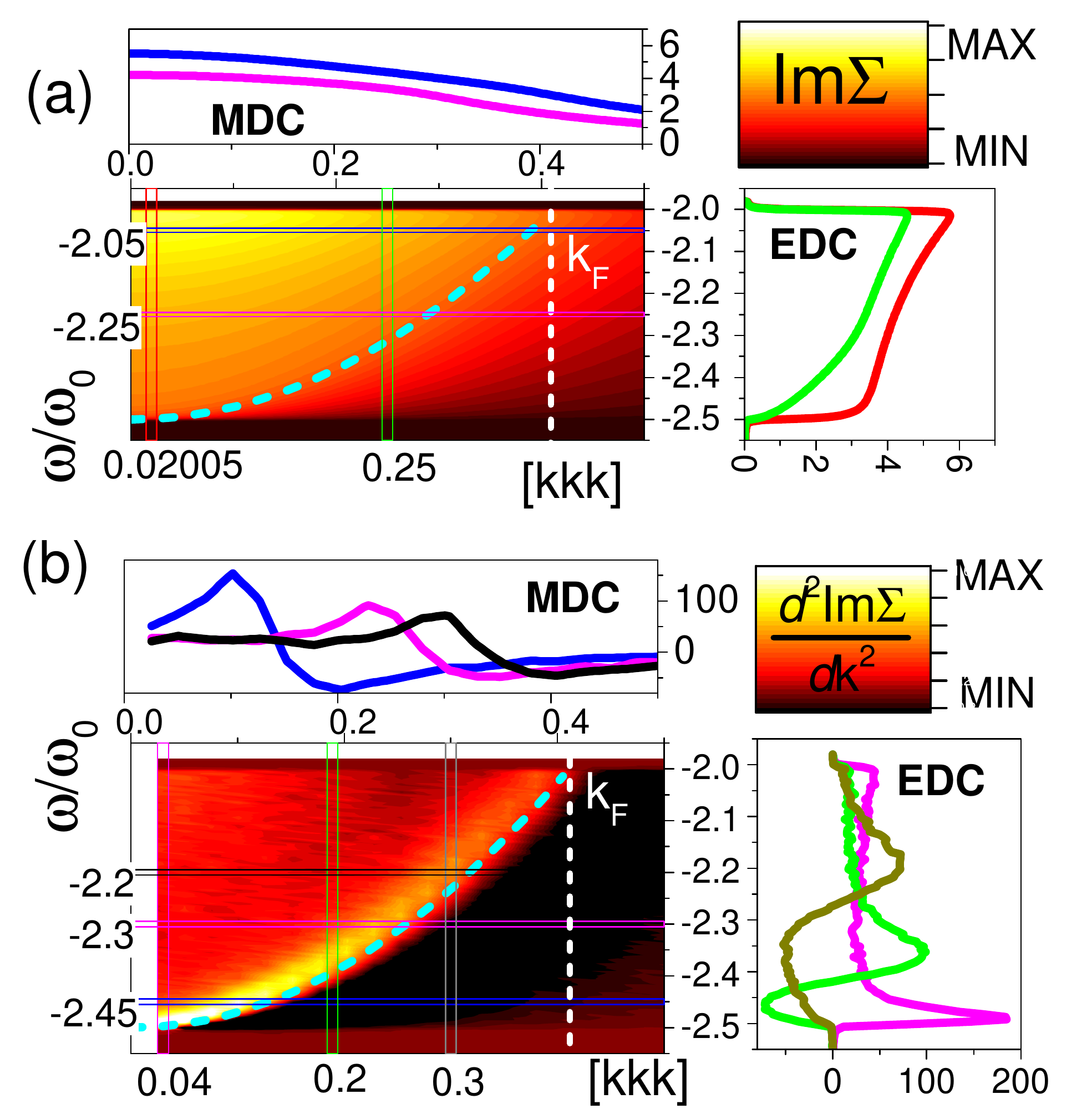}
		\caption{Contour plots of (a)
			$\mbox{Im}\Sigma^{(2)}({\bf k},\omega)$
			and (b) its second derivative $d^2  \mbox{Im}\Sigma^{(2)}({\bf k},\omega) / d \text{k}^2$
			for the 3D parameter set S1 and $r_{TF}=100$. 
			Right panels in (a) and (b) are the EDCs and upper panels are the MDCs 
			along the cuts shown in the left-bottom panels of (a) and (b). 
			The dashed curves represent the QP dispersion, shifted by 2$\omega_0$.}
		\label{fig:fig4}
	\end{center}
\end{figure}

In contrast to strong screenings, for weak screenings ($r_{TF}\gtrsim20$) the map of $\mbox{Im}\Sigma$ in Fig.~\ref{fig:fig3} exhibits maxima that almost exactly follow the QP dispersion. In fact, using this property the weak-screening regime can be unambiguously identified from the behavior of the first phonon sideband. On the other hand, in our results the EDC and momentum distribution curves (MDCs) maxima of $\mbox{Im}\Sigma$ for the second sideband are absent, with its typical intensity shown in Fig.~\ref{fig:fig4}a. By taking the second derivative $d^2 \mbox{Im}\Sigma^{(2)}({\bf k},\omega) / d \text{k}^2$, which is fully in line with the conventional experimental ARPES-data processing that uses the curvature analysis, one recovers a dispersion resembling to that of the QP peak. This is a general property of the second sideband, well-illustrated by Fig.~\ref{fig:fig4}, in which the intensities of $\mbox{Im}\Sigma^{(2)}({\bf k},\omega)$ and $d^2 \mbox{Im}\Sigma^{(2)}({\bf k},\omega) / d \text{k}^2$ are compared, together with the intensity dependence along the EDCs and the MDCs. We note that the necessity for the differential analysis of the second sideband follows purely from theoretical results, rather than being a consequence of a limited quality of particular experimental data.

\begin{figure}[tbp]
\begin{center}
\includegraphics[scale=0.3]{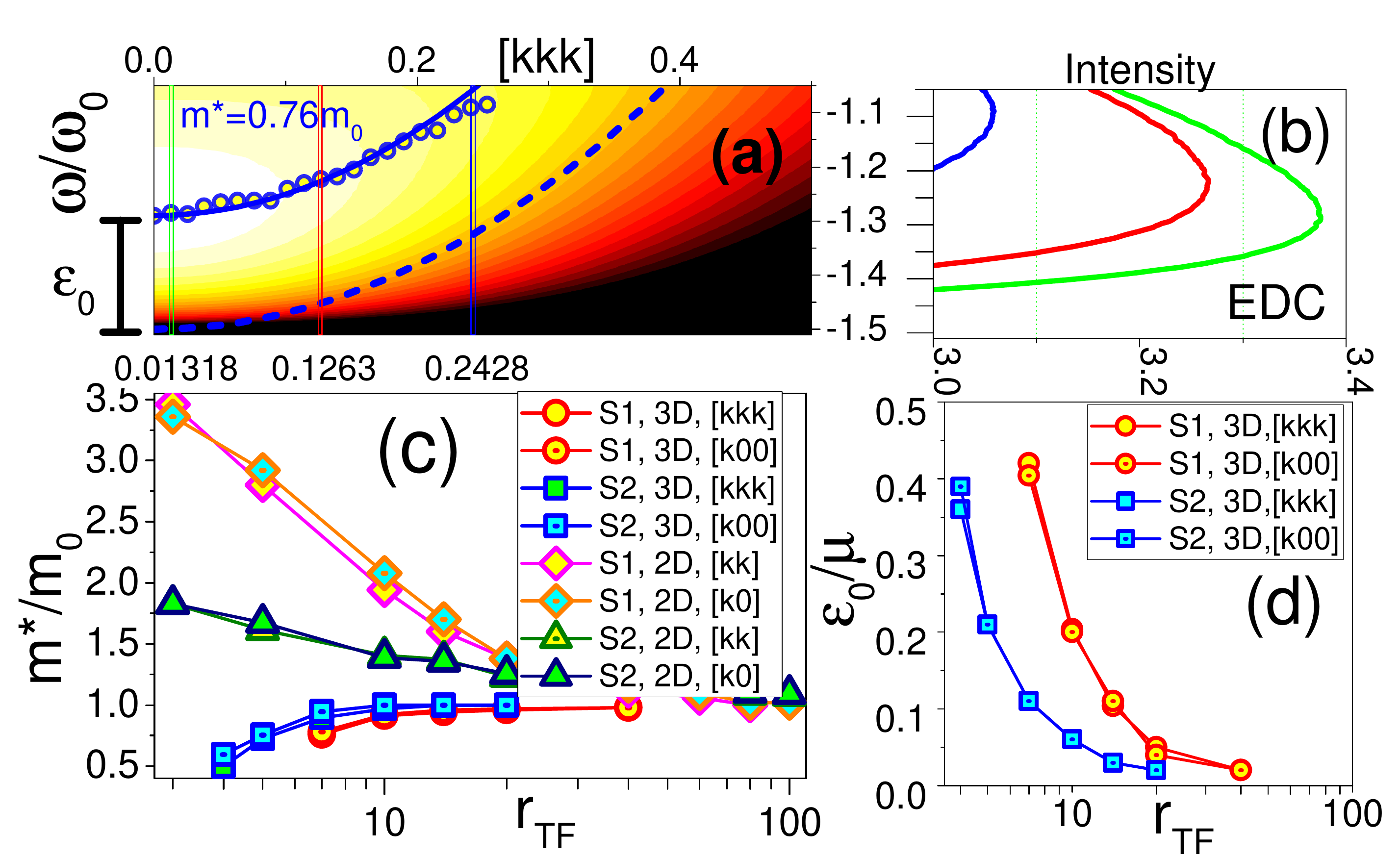}
\caption{(a) Contour plot of  
$\mbox{Im}\Sigma^{(1)}({\bf k},\omega)$ for the 3D parameter set S1 and 
$r_{TF}=7$. The dashed curve represents the QP dispersion $\epsilon_{\bf k}$, shifted downward by        
$\omega_0$. The circles follow the maxima of EDCs, fitted by a parabola (solid line) that corresponds to the mass $m^*=0.76\;m_0$. 
(b) EDCs for cross sections shown in panel (a). 
(c) $m^*$ derived from the first sideband for parameters when EDCs maxima do not follow the QP dispersion.
(d) Upward shift $\varepsilon_0$ characterizing the EDC maxima for the first phonon sideband.}
\label{fig:fig5}
\end{center}
\end{figure}

{\it Anomalous phonon sideband dispersions}. Between the strong and the weak screening regime, exhibiting clearly different and characteristic behaviors of $\mbox{Im}\Sigma$, another regime sets in, which can be recognized from anomalous phonon sideband dispersions. For intermediate EPI ranges, the structure of $\mbox{Im}\Sigma$ is very sensitive on parameters, which is illustrated by Fig.~\ref{fig:fig5}. The curve that follows the EDC maxima appears shifted upward by $\varepsilon_0$ from the lower edge of the first sideband. In comparison to the QP dispersion, parabolic fits of the EDCs maxima give different effective masses, e.g. for the $\mbox{Im}\Sigma$ intensity shown in Fig.~\ref{fig:fig5} one obtains $m^*=0.76\;m_0$. Such anomalous phonon sideband dispersions are found for all eight studied cases. Furthermore, anomalies in 3D and 2D are different: Figs.~\ref{fig:fig5}c-d show that $m^*$ is smaller (larger) in 3D (2D) and $\varepsilon_0$ varies (being zero) in 3D (2D). While in the 3D cases the EDC maxima follow the parabolic dispersion over the whole first sideband, in the 2D cases the parabolic dispersion at small momenta transforms, exhibiting a jump toward the large momenta.

\begin{figure}[bt]
\begin{center}
\includegraphics[scale=0.3]{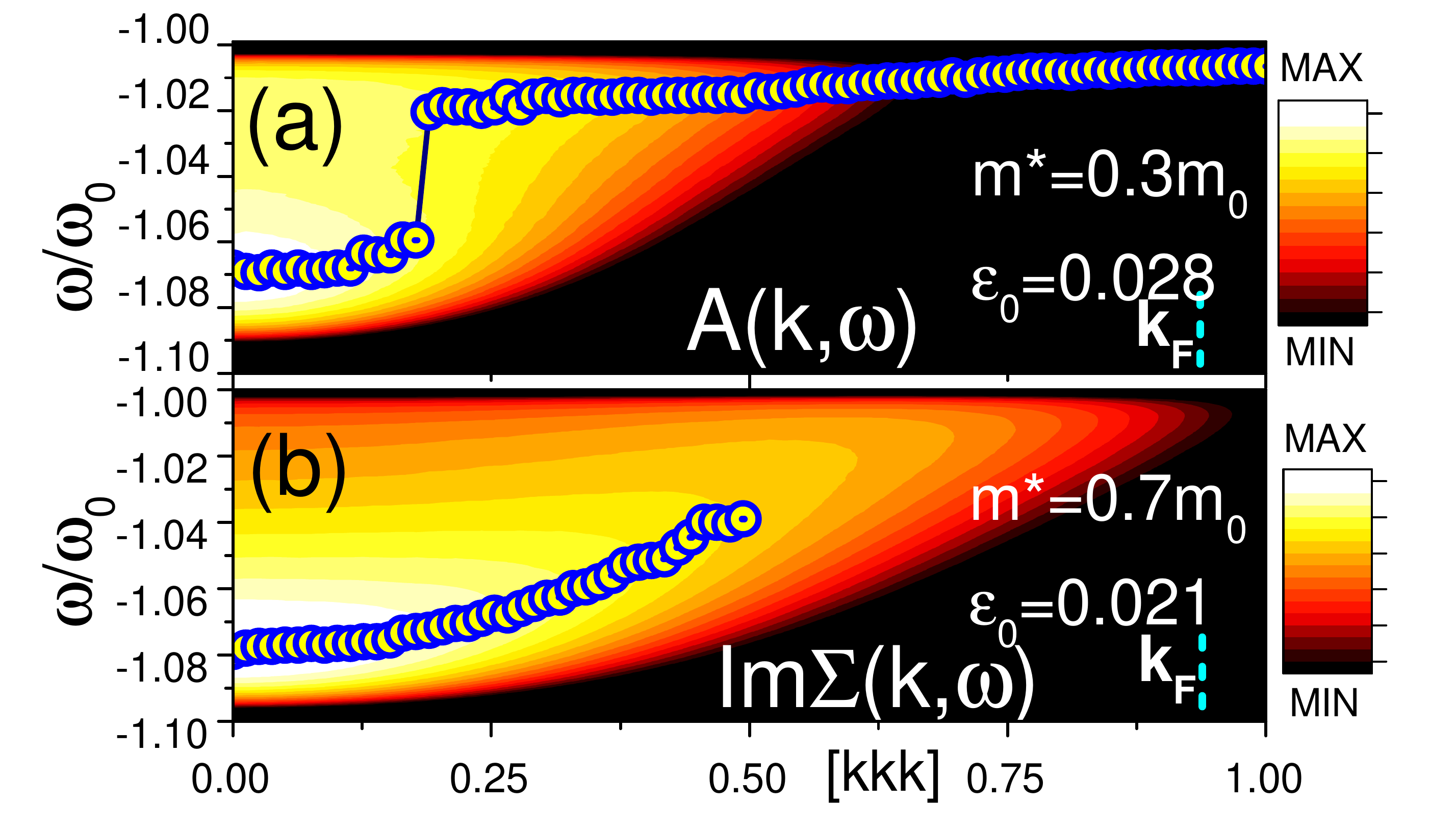}
\caption{Contour plot of the (a) ARPES component $A^{(1)}({\bf k},\omega)$
and (b) $\mbox{Im}\Sigma^{(1)}({\bf k},\omega)$ for the 3D parameter set S2 and $r_{TF}=5$. 
Circles follow maxima of EDCs. These are fitted by parabolas, giving corresponding effective masses $m^*$
and shifts $\varepsilon_0$ of the phonon sideband dispersion. }
\label{fig:fig6}
\end{center}
\end{figure}

The anomalous sideband dispersion, which can be obtained either from the EDCs or from the standard curvature analysis, characterizes the ARPES spectra as well. Importantly, as seen from Fig.~\ref{fig:fig6}a, the values of $\varepsilon_0$ and $m^*$ obtained from these spectra and from $\mbox{Im}\Sigma$ are different. 
For the intermediate screening length in Fig.~\ref{fig:fig6}, $r_{TF}=5$, the confinement of the 
ARPES intensity is enhanced in comparison to that obtained from  $\mbox{Im}\Sigma$, 
being reduced to momenta, $k<k_r$, where $k_r$ is considerably smaller than $k_F$. This effect is experimentally observed as well \cite{TiO2MoserBarisic2013,Cancellieri2016,Riley2018}. In general, for all the considered models and dimensionalities, we find that all anomalous behaviors are restricted to the specific screening range, $3 \le r_{TF}\le 20$.
 
{\it Confinement estimators.} Depending on the behavior of the first phonon sideband, our analysis identified three different regimes as a function of the range of the EPI. The calculations as well elucidated few features of ARPES sidebands that are independent of specific details of a measurement or material parameters, like the direction [klm]/[kl], the dimensionality 3D/2D or the parameter set S1/S2. In particular, the confinement of the ARPES sidebands in the Brillouin zone is a common property for all three screening regimes. On the other hand, the confinement of the $\mbox{Im}\Sigma$ intensity depends strongly on $r_{TF}$. In fact, a procedure based on this confinement may be developed to determine the range of the EPI from experimental ARPES spectra.

Let us start by considering the $\mbox{Im}\Sigma$ intensity integrated over the energies within the $n$-th phonon sideband for a given ${\bf k}$, ${\cal R}_n({\bf k})  = \int_{-n\omega_0-\mu}^{-n\omega_0} d\omega \mbox{Im}\Sigma^{(n)}({\bf k},\omega)$. Then, using ${\cal R}_{n}({\bf k})$, the confinement within the FS may be expressed by an estimator corresponding to the ratio of the intensity within and outside the FS,

\begin{equation}
R_n = 
\left(
\int_0^{k_F} dk \frac{{\cal R}_n({\bf k})}{k_F}
\right)
/
\left(
\int_{k_F}^{\pi} dk \frac{{\cal R}_n({\bf k})}{\pi-k_F}
\right) \; .
\label{RRR}
\end{equation}

\noindent Such estimator of the confinement is particularly suitable for an analysis of the experimental data since it involves averaging over frequency and momenta intervals, reducing effects of noise in experimental ARPES spectra. Moreover, it is unaffected by the arbitrariness in the normalization of the experimental ARPES intensity, as well as it is independent of the coupling strength $\alpha$. Furthermore, the regime of strong screening, when $\mbox{Im}\Sigma$ is ${\bf k}$-independent, may be identified directly from $R_{n}$, $R_{n}\rightarrow1$ when $r_{TF} \to 0$.

With the exception of the $r_{TF} \to 0$ limit, $R_{n}$ is insufficient to estimate $r_{TF}$. Namely, for different systems the maximal value of $R_n$ may change by an order of magnitude. Instead of $R_{n}$, thus, in such situations we consider an another estimator of the confinement,

\begin{equation} 
C_{n} (r_{TF}) = \left[ R_{n} (r_{TF})-1 \right] \left[ R_{n} (\infty)-1\right]^{-1}\;.
\label{Ce}
\end{equation}

\begin{figure}[tbp]
	\begin{center}
		\includegraphics[scale=0.3]{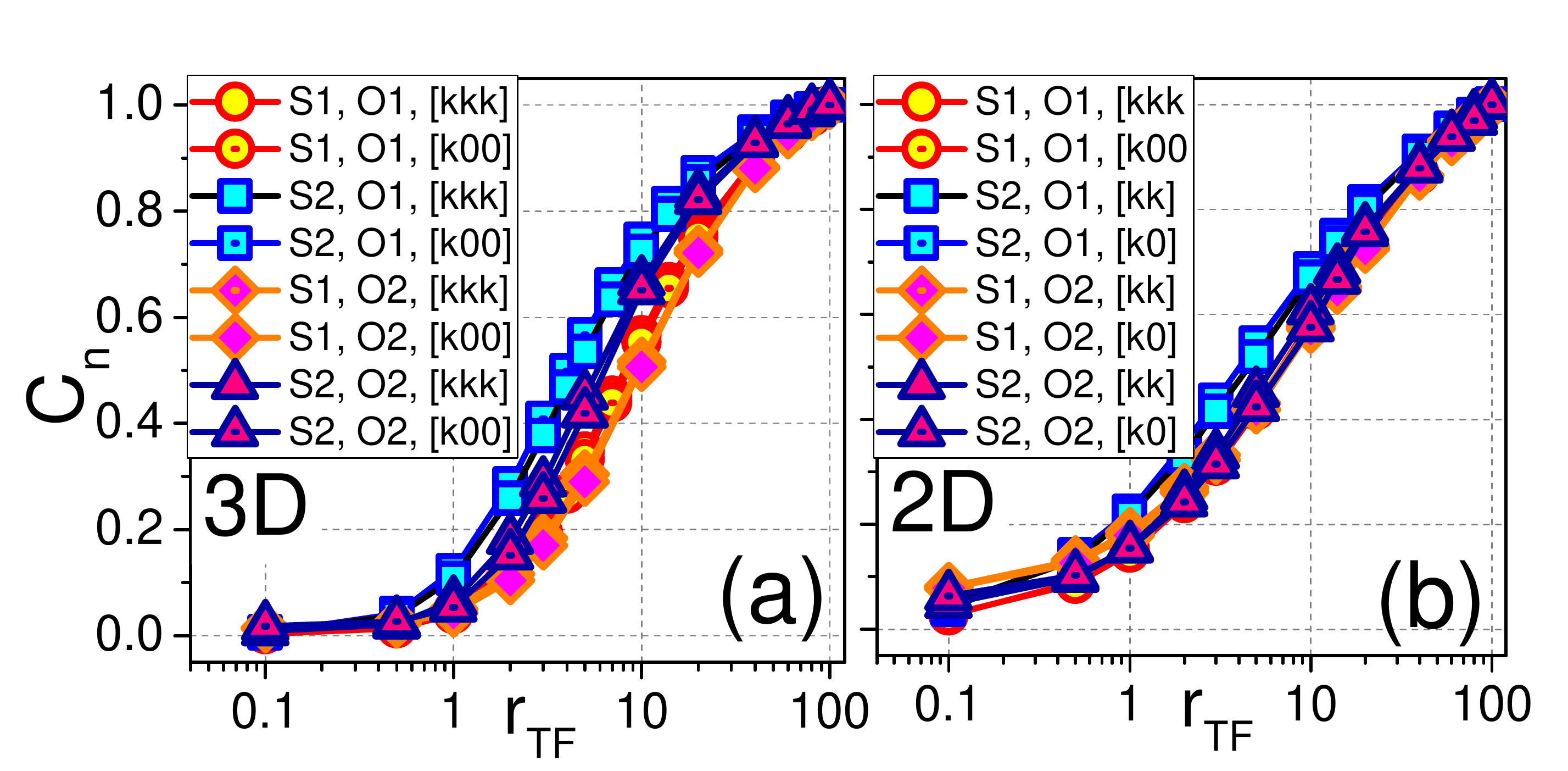} 
		\caption{Confinement parameter $C_{n}(r_{TF})$ (\ref{Ce}) for the parameter
			sets S1 and S2, the first O1 and the second O2 phonon sidebands in (a) 3D and (b) 2D.}
		\label{fig:fig2}
	\end{center}
\end{figure}

\noindent For all eight cases (parameter sets S1/S2, 3D/2D, diagonal $[k'k'k']/[k'k']$ and nondiagonal $[k'00]/[k',0]$ directions) and for the both sidebands, the estimator in Eq.~(\ref{Ce}) exhibits a fairly universal behavior, which is well illustrated by Fig.~\ref{fig:fig2}. $C_{n} (r_{TF})$ is scaled in Eq.~(\ref{Ce}) by the unscreened $r_{TF} \to \infty$ value of $R_{n}$. This value may be obtained from measurements of the reference material with unscreened EPI (small dopings), particularly for systems and experimental setups for which the charge density may easily be controlled. Alternatively, for the first sideband, $R_{1}(\infty)$ may easily be obtained numerically from closed expressions for $\mbox{Im}\Sigma^{(1)}({\bf k},\omega)$ for the 2D \cite{Supplement} and 3D \cite{Mahan} systems. Namely, all parameters that define $R_{1}(\infty)$ (effective mass $m^*$, Fermi level $\mu$ and $k_F$)  may be extracted directly from the experimental QP dispersion.

\begin{figure}[bt]
\begin{center}
\includegraphics[scale=0.21]{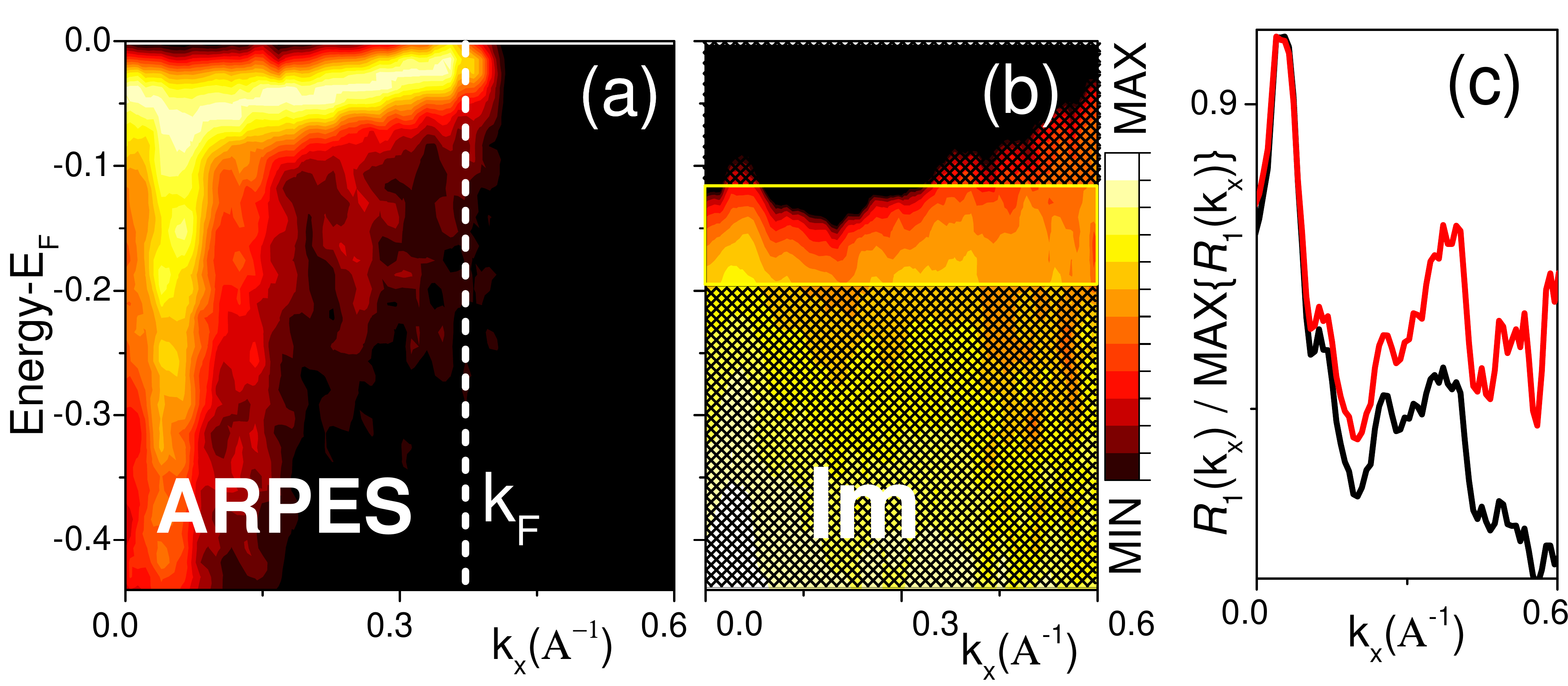}
\caption{(a) Raw ARPES data with the constant background removed. 
(b) $\mbox{Im}\Sigma$ corresponding to (a). (c) Measure of confinement
${\cal R}_1(k_x)/\mbox{MAX}\{{\cal R}_1(k_x)\}$ for data in (a) [black] and (b) [red] panel.\label{fig:fig7}}
\end{center}
\end{figure}

{\it Experimental data}. We end our analysis with an illustrative example, dealing with a particularly complex experimental realization of EPI effects for low electron densities. We consider the ARPES spectrum of oxygen-deficient LaAlO$_3$/SrTiO$_3$ interface \cite{Chikina2018}, involving different phonons and limited experimental resolution (0.04 eV), making a theoretical modeling difficult. Yet, even in such circumstances, our approach may provide valuable insights on the range of the EPI. Figure~\ref{fig:fig7}a shows details of the APRES spectrum, after the subtraction of the structureless background. $k_F\approx 0.37$ A$^{-1}$ and $\mu\approx0.07$ eV may easily be estimated from the QP dispersion.

As explained in the discussion that follows Eq.~(\ref{SF}), with QP properties known, one may obtain $\mbox{Im}\Sigma$ in Fig.~\ref{fig:fig7}b from the ARPES data in  Fig.~\ref{fig:fig7}a (for details see \cite{Supplement}). The frequency window in Fig.~\ref{fig:fig7}b dominated by the coupling to the LO3 phonon is highlighted separately, corresponding to the first LO3 sideband ($\omega_{LO3}\approx0.12$ eV \cite{Cancellieri2016}). By averaging ${\cal R}_1({\bf k})$ over these frequencies, $-\mu-\omega_{LO3}\leq \omega\leq-\omega_{LO3}$, one obtains $R_1 = 1.03$. This is almost the minimal value that the estimator $R_1$ can take, which alone clearly indicates the short-range EPI and the strong screening limit. The later, as analyised theoretically in Fig.~\ref{fig:fig1}, is characterized by the almost $k$-independent $\mbox{Im}\Sigma$. For a more complicated structure of $\mbox{Im}\Sigma$, further considerations based on $C_1(r_{TF})$ in Eq.~(\ref{Ce}) would be required. We are stressing that the confinement of ARPES spectral weight in Fig.~\ref{fig:fig7}a is much stronger ($R_1 = 1.85$), because of which one may incorrectly conclude about the long-range character of the coupling to the LO3 phonons.

{\it Conclusions}. Our study provides means to estimate the range/screening length of the EPI directly from the ARPES spectra, giving important insights into the polaronic correlations present in a wide class of real systems involving many puzzling physical phenomena. The ${\bf k}$- and $\omega$-dependencies of the first and the second sideband are analyzed in details as a function of screening, in terms of exact leading corrections obtained from the diagrammatic expansion, including the leading vertex correction. It is shown that the range of the EPI may be extracted from the confinement of $\mbox{Im}\Sigma$ within the Brillouin zone, even when the experimental resolution is very limited. Our results apply to all recently investigated low-density metallic systems, characterized by the screened Fr\" ohlich interaction, while our methodology is not restricted to phonons only, since coupling to excitations of other nature (plasmons, magnons, paramagnons, charge-order fluctuations, etc.) may be analyzed in a similar manner.

{\em Acknowledgements.} N.N. and A.S.M acknowledge the support by JST CREST Grant Number JPMJCR1874, Japan.
J.K acknowledges the support by the Croatian Science Foundation Project IP-2016-06-7258. 
O.S.B. acknowledges the support by the QuantiXLie Center of Excellence, a project co-financed by the Croatian Government and European Union (Grant KK.01.1.1.01.0004).

\bibliography{paper}{}

\end{document}